\documentclass{ws-ijgmmp}
\usepackage{amssymb,amsmath,latexsym,mathrsfs,csquotes,blkarray}
\newtheorem{thm}{Theorem}[section]

 \newtheorem{prop}[thm]{Proposition}
 \newtheorem{defn}[thm]{Definition}
\def\nmspace{\mathbb R^{n|m}}
\def\functions{C^{\infty}(\mathbb R^{n|m})}
\def\evenfunctions{C^{\infty}_0(\mathbb R^{n|m})}

\begin{document}
\title{Nambu-Poisson bracket on superspace}
\author{VIKTOR ABRAMOV}
\address{Institute of Mathematics and Statistics, University of Tartu, J. Liivi 2 - 602\\
Tartu 50409, Estonia\\
\email{viktor.abramov@ut.ee}}
\maketitle
\begin{history}
\received{(May 2018)}
\revised{(August 2018)}
\end{history}
\begin{abstract}
{We propose an extension of $n$-ary Nambu-Poisson bracket to superspace $\mathbb R^{n|m}$ and construct by means of superdeterminant a family of Nambu-Poisson algebras of even degree functions, where the parameter of this family is an invertible transformation of Grassmann coordinates in superspace $\mathbb R^{n|m}$. We prove in the case of the superspaces $\mathbb R^{n|1}$ and $\mathbb R^{n|2}$ that our $n$-ary bracket, defined with the help of superdeterminant, satisfies the conditions for $n$-ary Nambu-Poisson bracket, i.e. it is totally skew-symmetric and it satisfies the Leibniz rule and the Filippov-Jacobi identity (fundamental identity). We study the structure of $n$-ary bracket defined with the help of superdeterminant in the case of superspace $\mathbb R^{n|2}$ and show that it is the sum of usual $n$-ary Nambu-Poisson bracket and a new $n$-ary bracket, which we call $\chi$-bracket, where $\chi$ is the product of two odd degree smooth functions.}
\end{abstract}
\section{Introduction}
A generalization of Hamiltonian mechanics by means of a ternary (or, more generally, $n$-ary) bracket of functions determined on a phase space was proposed by Y. Nambu in \cite{Nambu} and developed in the series of papers. An excellent introduction to this field of research is given in \cite{Takhtajan}. The peculiar property of generalization of Hamiltonian mechanics proposed by Y. Nambu is that the Nambu-Hamilton equation of motion, which describes the dynamics, contains $n-1$ Hamilton functions. It is mentioned in \cite{Takhtajan} that Y. Nambu proposed and developed his generalization of Hamiltonian mechanics based on a notion of triple bracket with the goal to apply this approach to quarks model, where baryons are combinations of three quarks. Independently of Y. Nambu, V.T. Filippov proposed a notion of $n$-Lie algebra, which is a generalization of the notion of Lie algebra based on $n$-ary Lie bracket \cite{Filippov}. The basic component of the definition of $n$-Lie algebra is generalized Jacobi identity, which is now called either fundamental identity or Filippov-Jacobi identity. Later it turned out that a generalization of Hamiltonian mechanics proposed by Y. Nambu and a generalization of Lie algebra proposed by V.T. Filippov are closely related. Particularly, it was shown that ternary Nambu bracket (or, more generally, $n$-ary bracket) satisfies the Filippov-Jacobi identity. The question of quantization of Nambu-Poisson bracket has been considered in a number of papers, but so far this is the outstanding problem. In the paper \cite{Awata-Li-Minic-Yaneya} the authors propose the realization of quantum Nambu-Poisson bracket by means of $n$th order matrices, where the triple commutator is defined with the help of usual commutator and the trace of a matrix. This approach is extended to super Nambu-Poisson bracket by means of supermatrices, where the $\mathbb Z_2$-graded triple commutator is defined with the help of the supertrace of a supermatrix \cite{Abramov 1}, \cite{Abramov 2}.

\vskip.3cm
\noindent
Let us remind a concept of Poisson manifold. Let $M$ be a smooth finite dimensional manifold and $\frak C=C^\infty(M)$ be the algebra of smooth functions on this manifold. A bilinear mapping $\{\cdot,\cdot\}:\frak C\times\frak C\to \frak C$ is said to be a Poisson bracket if for any smooth functions $f,g,h\in\frak C$ it satisfies
\begin{itemize}
\item[i)]
$\{f,g\}=-\{g,f\}$ (skew-symmetry);
\item[ii)]
$\{f\,g,h\}=f\;\{g,h\}+g\;\{f,h\}$ (Leibniz rule);
\item[iii)]
$\{f,\{g,h\}\}+\{g,\{h,f\}\}+\{h,\{f,g\}\}=0.$ (Jacobi identity).
\end{itemize}
A smooth manifold $M$ endowed with a Poisson bracket is referred to as a Poisson manifold. For instant consider the 2-dimensional space $\mathbb R^2$ with coordinates denoted by $p,q$ and define the binary bracket by the formula
\begin{equation}
\{f,g\}=\mbox{Det}\,\left(
                      \begin{array}{cc}
                        \partial_p f & \partial_q f\\
                        \partial_p g& \partial_q g\\
                      \end{array}
                    \right),
\label{Poisson 2-dimensional}
\end{equation}
where $\partial_p f=\frac{\partial f}{\partial p},\partial_q f=\frac{\partial f}{\partial q}$.
Then it is easy to show that (\ref{Poisson 2-dimensional}) is the Poisson bracket. Hence the 2-dimensional space $\mathbb R^2$ endowed with the binary bracket (\ref{Poisson 2-dimensional}) is the Poisson manifold.

\vskip.3cm
\noindent
A generalization of Poisson bracket was proposed by Y. Nambu in \cite{Nambu}, where he introduced a ternary bracket of three smooth functions $f,g,h$ defined on the three dimensional space $\mathbb R^3$, whose coordinates are denoted by $x,y,z$. This ternary bracket is defined with the help of the Jacobian of a mapping
$$
(x,y,z)\to (f(x,y,z),g(x,y,z),h(x,y,z))
$$
as follows
\begin{equation}
\{f,g,h\}=\frac{\partial(f,g,h)}{\partial(x,y,z)}=\mbox{Det}\,\left(
                                                                \begin{array}{ccc}
                                                                  \partial_x f & \partial_y f & \partial_z f \\
                                                                  \partial_x g & \partial_y g & \partial_z g \\
                                                                  \partial_x h & \partial_y h & \partial_z h \\
                                                                \end{array}
                                                              \right).
\label{Nambu ternary bracket}
\end{equation}
Evidently this ternary bracket is totally skew-symmetric. It can be also verified that it satisfies the Leibniz rule
$$
\{g\,h,f^1,f^2\}=g\;\{h,f^1,f^2\}+h\;\{g,f^1,f^2\},
$$
and the identity
\begin{equation}
\{g,h,\{f^1,f^2,f^3\}\}=\{\{g,h,f^1\},f^2,f^3\}+\{f^1,\{g,h,f^2\},f^3\}+\{f^1,f^2,\{g,h,f^3\}\}.\nonumber
\end{equation}
This identity is called either fundamental identity or Filippov-Jacobi identity and its $n$-ary version is the basic component of a concept of $n$-Lie algebra proposed by V.T. Filippov in \cite{Filippov}.

\vskip.3cm
\noindent
The ternary Nambu bracket (\ref{Nambu ternary bracket}) can be generalized to any number of arguments as follows. Let $\frak C$ be the algebra of smooth functions on a smooth finite dimensional manifold $M$. Then
a multilinear mapping $\{\cdot,\cdot,\ldots,\cdot\}:\frak C\times\frak C\times\ldots\times\frak C\; (n\;\mbox{times})\to\frak C$ is called a $n$-ary Nambu-Poisson bracket if for any smooth functions $f^1,f^2,\ldots,f^n$, $g^1,g^2,\ldots,g^n$ it is totally skew-symmetric, satisfies the Leibniz rule
$$
\{g^1\, g^2,f^1,\ldots,f^{n-1}\}=g^1\,\{g^2,f^1,\ldots,f^{n-1}\}+g^2\,\{g^1,f^1,\ldots,f^{n-1}\},
$$
and the Filippov-Jacobi identity
\begin{eqnarray}
\{f^1,\ldots,f^{n-1},\{g^1,g^2,&\ldots&,g^n\}\}=\nonumber\\
       && \sum_{\mu=1}^n\,\{g^1,\ldots,g^{\mu-1},\{f^1,\ldots,f^{n-1},g^\mu\},g^{\mu+1},\ldots,g^n\}.\nonumber
\end{eqnarray}
A smooth manifold $M$ endowed with a $n$-ary Nambu-Poisson bracket is called a Nambu-Poisson manifold of $n$th order \cite{Takhtajan}.

\vskip.3cm
\noindent
The aim of this paper is to extend the concept of Nambu-Poisson manifold to supermanifolds. By supermanifold we mean a ringed space of certain type as it was proposed by Berezin \cite{Berezin}. It is worth to briefly remind the approach of Berezin. Let $\frak G^m$ be a Grassmann algebra with $m$ generators $\theta^1,\theta^2,\ldots,\theta^m$. Then we associate to each open subset $U\subset \mathbb R^n$ the ring of smooth $\frak G^m$-valued functions $C^\infty(U;\frak G^m)$, where $U$ is a domain of functions. This defines the sheaf of rings on the $n$-dimensional space $\mathbb R^n$, where a structure morphisms $\rho^U_V:C^\infty(U;\frak G^m)\to C^\infty(V;\frak G^m)$ ($V\subset U$) is simply the restriction of a function to a subset, i.e. $\rho^U_V(f)=f|_V, f\in C^\infty(U;\frak G^m)$. Then the $n$-dimensional space $\mathbb R^n$ endowed with the sheaf of rings of smooth $\frak G^m$-valued functions is the ringed space which we will denote by $\mathbb R^{n|m}$ and this ringed space is referred to as the superspace. This superspace serves as a model space for a supermanifold. This means that a supermanifold $M^{n|m}$ locally in a neighborhood  of each point looks like $\mathbb R^{n|m}$. More precisely, a supermanifold $M^{n|m}$ is a smooth manifold $M^n$ endowed with a sheaf of rings of smooth $\frak G^m$-valued functions, where $\frak G^m$ is a Grassmann algebra with $m$ generators. In a neighborhood $U$ of a point $p\in M^n$ we can introduce a set of even coordinates $x^\mu$ and odd coordinates $\theta^\alpha$. For two neighborhoods $U,V$ ($U\cap V\neq\emptyset$) with local coordinates $x^\mu,\theta^\alpha$ and $y^\nu,\eta^\beta$ respectively we have the transition functions (from one local coordinate system to another)
\begin{equation}
x^\mu=x^\mu(y,\eta), \;\;\theta^\alpha=\theta^\alpha(y,\eta),
\label{transformation of coordinates in superspace}
\end{equation}
where $x^\mu(y,\eta)$ are even degree functions and $\theta^\alpha(y,\eta)$ are odd degree functions. It should be mentioned that according to definition of a supermanifold the mapping $x^\mu=x^\mu(y,\eta)|_{\eta=0}$ induced by a transformation of local coordinates (\ref{transformation of coordinates in superspace}) must determine the diffeomorphism of two open subsets of $\mathbb R^n$.

\vskip.3cm
\noindent
The superalgebra of smooth functions $C^\infty(M^{n|m})$ on a supermanifold $M^{n|m}$ splits into direct sum of subalgebra of even degree functions $C_0^\infty(M^{n|m})$ and the subspace of odd degree functions $C_1^\infty(M^{n|m})$. In analogy with Nambu-Poisson manifold we give the definition
\begin{defn}
A supermanifold $M^{n|m}$ endowed with a $n$-ary bracket of $n$ even degree smooth functions, which is totally skew-symmetric, satisfies the Leibniz rule and the Filippov-Jacobi identity, is referred to as Nambu-Poisson supermanifold of $n$th order.
\end{defn}
In this paper we construct a $n$-ary bracket of even degree smooth functions defined on a superspace $\mathbb R^{n|m}$. In analogy with approach proposed by Y. Nambu we define a $n$-ary bracket of even degree smooth functions by means of superdeterminant of supermatrix, whose entries are the derivatives of functions with respect to coordinates of superspace $\mathbb R^{n|m}$. More precisely, if we consider a superspace $\mathbb R^{n|m}$ then an analog of Jacobian of a diffeomorphism is the Berezinian of a transformation (\ref{transformation of coordinates in superspace}). We consider the odd part of  transformation (\ref{transformation of coordinates in superspace}), i.e. the collection of $m$ odd degree functions $\Psi=(\theta^\alpha(y,\eta))_{\alpha=1}^m$, as the parameter of $n$-ary bracket. In Section 3 and 4 we prove in the case of the superspace with one and two Grassmann coordinates that our $n$-ary bracket satisfies the condition of total skew-symmetry, the Leibniz rule and the Filippov-Jacobi identity. Hence we prove that the superspaces $\mathbb R^{n|1}$ and $\mathbb R^{n|2}$ endowed with a $n$-ary bracket, constructed by means of superdeterminant, are Nambu-Poisson supermanifolds. We also show that $n$-ary bracket, constructed with the help of superdeterminant, is the sum of usual $n$-ary Nambu-Poisson bracket and the new $n$-ary bracket, which we call $\chi$-bracket, where $\chi$ is the product of two odd degree functions, which are the components of $\Psi$. We prove in the case of the superspace $\mathbb R^{n|2}$ that $n$-ary $\chi$-bracket also satisfies the conditions of total skew-symmetry, the Leibniz rule and the Filippov-Jacobi identity. Hence it also determines the structure of Nambu-Poisson supermanifold on the superspace $\mathbb R^{n|2}$.
\section{Superdeterminant and $n$-ary bracket in superspace $\mathbb R^{n|m}$}
In this section we define the $n$-ary bracket of $n$ even degree smooth functions defined on a superspace $\mathbb R^{n|m}$ by means of superdeterminant. Here we follow an analogy with the approach of Y. Nambu, where the Jacobian of a diffeomorphism of $n$-dimensional space $\mathbb R^n$ is used to define a $n$-ary analog of Poisson bracket. In the case of superspace $\mathbb R^{n|m}$ with coordinates $x^\mu,\theta^a$, where $\mu=1,2,\ldots,n$ and $a=1,2,\ldots,m$, a transformation of coordinates has the form
$$
x^\mu=x^\mu(y,\eta),\;\;\;\theta^a=\theta^a(y,\eta),
$$
where $y^\nu,\eta^b$ are new coordinates of superspace $\mathbb R^{n|m}$, $x^\mu(y,\eta)$ are even degree smooth functions and $\theta^a(y,\eta)$ are odd degree smooth functions. We can consider the Berezinian of this transformation, which is the analog of Jacobian. Here it is important that, in spite of the fact that the Berezinian includes derivatives of odd degree functions (with respect to coordinates of superspace), its value is always the even degree function. Therefore, we use the superdeterminant to define a $n$-ary bracket of $n$ even degree smooth functions $x^\mu(y,\eta)$, treating $m$ odd degree smooth functions $\theta^a(y,\eta)$ as parameters of $n$-ary bracket.

\vskip.3cm
\noindent
Let $\mathbb R^{n|m}$ be the superspace with $n$ real (even degree) coordinates $x^\mu$ and $m$ anticommuting (odd degree) coordinates $\theta^a$. The coordinates $\theta^a$ generate the Grassmann algebra and in order to expand any element of this algebra in terms of generators $\theta^a$ we will use the following notations: Let $\mathscr M=\{1,2,\ldots,m\}$ be the set of positive first $m$ integers, $I=\{a_1,a_2,\ldots,a_k\}$ be a subset of $\mathscr M$ and $\theta^I=\theta^{a_1}\theta^{a_2}\ldots\theta^{a_k},\; \theta^{\emptyset}=1$. Then any element of the Grassmann algebra generated by coordinates $\theta^a$ can be written in the form
$$
\sum_{I\subset \mathscr M}\lambda_I\,\theta^I,\quad \lambda_I=\lambda_{a_1a_2\ldots a_k}\in \mathbb R.
$$
In what follows we will denote the first term of this expansion $\lambda_{\empty}$ by $\lambda_0$.
Let $C^{\infty}(\mathbb R^{n|m})$ be the superalgebra of smooth functions on the superspace $\mathbb R^{n|m}$. Denote by $r$ the collection of real coordinates and by $\theta$ the collection of Grassmann coordinates, i.e. $r=(x^1,x^2,\ldots,x^n),\,\theta=(\theta^1,\theta^2,\ldots,\theta^m)$. Each smooth function $f(x,\theta)\in C^{\infty}(\mathbb R^{n|m})$ can be expanded in Grassmann coordinates as follows
\begin{equation}
f(r,\theta)=\sum_{I\subset\mathscr M}f_I(r)\,\theta^I,
\label{expansion of function in coordinates}
\end{equation}
where $f_{I}(r)=f_{a_1a_2\ldots a_k}(r)\in C^\infty(\mathbb R^n)$ is a smooth function.
Then
$$
C^{\infty}(\mathbb R^{n|m})=C^{\infty}_0(\mathbb R^{n|m})\oplus C^{\infty}_1(\mathbb R^{n|m}),
$$
where $C^{\infty}_0(\mathbb R^{n|m})$ is the subalgebra of even degree functions and $C^{\infty}_1(\mathbb R^{n|m})$ is the subspace of odd degree functions.

\vskip.3cm
\noindent
Now let
\begin{equation}
y^\mu=f^\mu(r,\theta),\;\;\eta^a=\psi^a(r,\theta),
\label{diffeomorphism}
\end{equation}
where $f^\mu(r,\theta),\mu=1,2,\ldots,n$ are even degree functions and $\psi^a(r,\theta),a=1,2,\ldots,m$ are odd degree functions, be a diffeomorphism of the superspace $\nmspace$. This diffeomorphism can be split into even degree part $y^\mu=f^\mu(r,\theta)$ and the odd degree part $\eta^a=\psi^a(r,\theta)$, which we denote by $\Psi$. Thus
$$
\Psi: \eta^a=\psi^a(r,\theta).
$$
Let us denote by $\frak B$ the Berezin supermatrix of a diffeomorphism (\ref{diffeomorphism}). Then $\frak B$ can be written as block matrix
\begin{equation}
\frak B=\left(
    \begin{array}{cc}
      \frak B_{11} & \frak B_{12} \\
      \frak B_{21} & \frak B_{22} \\
    \end{array}
  \right),
  \label{supermatrix}
\end{equation}
where $\frak B_{11},\frak B_{22}$ are even degree blocks
\begin{eqnarray}
\frak B_{11}=\left(
             \begin{array}{cccc}
               \partial_{x^1}f^1 & \partial_{x^2}f^1 & \ldots & \partial_{x^n}f^1 \\
               \partial_{x^1}f^2 & \partial_{x^2}f^2 & \ldots & \partial_{x^n}f^2 \\
               \ldots & \ldots & \ldots & \ldots \\
               \partial_{x^1}f^n & \partial_{x^2}f^n & \ldots & \partial_{x^n}f^n \\
             \end{array}
           \right),
           \frak B_{22}=\left(
             \begin{array}{cccc}
               \partial_{\theta^1}\psi^1 & \partial_{\theta^2}\psi^1 & \ldots & \partial_{\theta^m}\psi^1 \\
               \partial_{\theta^1}\psi^2 & \partial_{\theta^2}\psi^2 & \ldots & \partial_{\theta^m}\psi^2 \\
               \ldots & \ldots & \ldots & \ldots \\
               \partial_{\theta^1}\psi^m & \partial_{\theta^2}\psi^m & \ldots & \partial_{\theta^m}\psi^m \\
             \end{array}
           \right),
\nonumber
\end{eqnarray}
and $\frak B_{12}, \frak B_{21}$ are odd degree blocks
\begin{equation}
\frak B_{12}=\left(
             \begin{array}{cccc}
               \partial_{\theta^1}f^1 & \partial_{\theta^2}f^1 & \ldots & \partial_{\theta^m}f^1 \\
               \partial_{\theta^1}f^2 & \partial_{\theta^2}f^2 & \ldots & \partial_{\theta^m}f^2 \\
               \ldots & \ldots & \ldots & \ldots \\
               \partial_{\theta^1}f^n & \partial_{\theta^2}f^n & \ldots & \partial_{\theta^m}f^n \\
             \end{array}
           \right),
           \frak B_{21}=\left(
             \begin{array}{cccc}
               \partial_{x^1}\psi^1 & \partial_{x^2}\psi^1 & \ldots & \partial_{x^n}\psi^1 \\
               \partial_{x^1}\psi^2 & \partial_{x^2}\psi^2 & \ldots & \partial_{x^n}\psi^2 \\
               \ldots & \ldots & \ldots & \ldots \\
               \partial_{x^1}\psi^m & \partial_{x^2}\psi^m & \ldots & \partial_{x^n}\psi^m \\
             \end{array}
           \right),
\nonumber
\end{equation}
The superdeterminant of the Berezin supermatrix $\frak B$ \cite{Berezin}, which is also called the Berezinian, is defined by
\begin{equation}
\mbox{Sdet}\,\frak B=\mbox{Det}(\frak B_{11}-\frak B_{12}\,\frak B_{22}^{-1}\,\frak B_{21})\;\mbox{Det}\,\frak B_{22}^{-1}.
\label{berezinian}
\end{equation}
Since we assume that (\ref{diffeomorphism}) is a diffeomorphism of the superspace $\mathbb R^{n|m}$, the matrix $\frak B_{22}$ is invertible and the determinant of $\frak B_{22}$ will be denoted by $\Delta$, i.e. $\Delta=\mbox{Det}\,\frak B_{22}\neq 0$.

\vskip.3cm
\noindent
Now our aim is to extend a Nambu-Poisson $n$-ary bracket \cite{Nambu,Takhtajan} to functions on the superspace $\mathbb R^{n|m}$ by means of superdeterminant (\ref{berezinian}). The elements of the matrix $\frak B_{11}-\frak B_{12}\,\frak B_{22}^{-1}\,\frak B_{21}$ are even degree functions, the value of determinant of $\frak B^{-1}_{22}$ is also the even degree function. Consequently the value of the superdeterminant (\ref{berezinian}) is the even degree function. Thus if our aim is to construct an analog of Nambu-Poisson $n$-ary bracket for functions on the superspace $\mathbb R^{n|m}$ with the help of the superdeterminant, then we can do this only for even degree functions.

\vskip.3cm
\noindent
Let $f^1(r,\theta),f^2(r,\theta),\dots,f^n(r,\theta)$ be arbitrary even degree smooth functions on the superspace $\nmspace$, i.e. we do not assume that they determine an even part of a diffeomorphism (\ref{diffeomorphism}). Next we fix the odd part $\Psi$ of a diffeomorphism (\ref{diffeomorphism}), i.e. we choose $m$ odd degree functions $\psi^a(r,\theta)$ such that the matrix $\frak B_{22}$ is invertible at any point $r\in \mathbb R^n$. The collection of even degree functions $f^\mu(r,\theta)$ and odd degree functions $\psi^a(r,\theta)$ determines the supermatrix $\frak B$ (\ref{supermatrix}). In analogy with Nambu-Poisson $n$-bracket on a $n$-dimensional space (or manifold) \cite{Nambu},\cite{Takhtajan} we define the  $n$-ary bracket (which depends on a choice of collection $\Psi$ of odd degree functions) for $n$ even degree smooth functions $f_1(r,\theta)$,$f_2(r,\theta)$,$\dots$,$f_n(r,\theta)$ by the formula
\begin{equation}
\{f^1(r,\theta),f^2(r,\theta),\dots,f^n(r,\theta\}_{\Psi}=\mbox{Sdet}\,\frak B.
\label{definition of Nambu-Poisson n-bracket}
\end{equation}
In what follows we will call this $n$-ary bracket $n$-ary $\Psi$-bracket. It is worth to remind that $\Psi$ denotes the collection of odd degree smooth functions $\psi^a(r,\theta), a=1,2,\ldots,m$ such that the $m$th order matrix $\frak B_{22}$ is invertible, and the notation for $n$-ary bracket (\ref{definition of Nambu-Poisson n-bracket}) clearly shows its dependence on a choice of $\Psi.$ Thus we consider $\Psi=(\psi^a(r,\theta))_{a=1}^m$ as a parameter of $n$-ary $\Psi$-bracket (\ref{definition of Nambu-Poisson n-bracket}). It is worth to mention that $\Psi$ can be considered as the transformation $\eta^a=\psi^a(r,\theta)$ of Grassmann coordinates $\theta^a$ in the superspace $\nmspace$ and the set of all these transformations form the infinite-dimensional group, where the group operation is the product of two transformations. Hence we can assign to each element of this infinite-dimensional group the $n$-ary $\Psi$-bracket (\ref{definition of Nambu-Poisson n-bracket}) on the superspace $\mathbb R^{n|m}$.

\vskip.3cm
\noindent
It follows from the definition of $n$-ary $\Psi$-bracket (\ref{definition of Nambu-Poisson n-bracket}) that it determines the multilinear mapping
\begin{equation}
\{\cdot,\cdot,\ldots,\cdot\}_{\Psi}:\evenfunctions\times\ldots\times\evenfunctions\,(n\;\mbox{times})\to\evenfunctions,
\label{multilinear mapping}
\end{equation}
i.e. the $n$-ary $\Psi$-bracket (\ref{definition of Nambu-Poisson n-bracket}) is defined on the even subalgebra $\evenfunctions$ of the superalgebra $\functions$. It follows immediately from the definition of $n$-ary $\Psi$-bracket (\ref{definition of Nambu-Poisson n-bracket}) and from the properties of superdeterminant that the mapping (\ref{multilinear mapping}) is totally skew-symmetric, that is, if we perform a permutation of even degree functions in the $n$-ary $\Psi$-bracket (\ref{definition of Nambu-Poisson n-bracket}) then the $n$-ary bracket changes the sign in accordance with the parity of permutation.

\vskip.3cm
\noindent
Particularly assume $\Psi$ is the identity transformation of Grassmann coordinates $\theta^a$, i.e. $\eta^a=\psi^a(r,\theta)=\theta^a$, and functions $f^\mu(r), \mu=1,2\ldots,n$ do not depend on Grassmann coordinates. Then $\frak B_{22}$ is the unit matrix of $m$th order, $\frak B_{12},\frak B_{21}$ are zero matrices of dimensions $n\times m$ and $m\times n$ respectively. Thus $\mbox{Det}\,(\frak B^{-1}_{22})=1$, the product of matrices $\frak B_{12}\,\frak B_{22}^{-1}\,\frak B_{21}$ vanishes and from (\ref{berezinian}), (\ref{definition of Nambu-Poisson n-bracket}) it follows that in this case the $n$-ary $\Psi$-bracket (\ref{definition of Nambu-Poisson n-bracket}) reduces to usual $n$-ary Nambu-Poisson bracket, i.e.
$$
\{f^1(r),f^2(r),\dots,f^n(r)\}_{\Psi}=\mbox{Det}\,(\partial_{x^\mu}f^\nu(r))=\{f^1(r),f^2(r),\ldots,f^n(r)\},
$$
where $\{f^1(r),f^2(r),\ldots,f^n(r)\}$ is the usual Nambu-Poisson $n$-bracket for functions on the space $\mathbb R^n$. Thus we can consider the $n$-ary $\Psi$-bracket as an extension of the $n$-ary Nambu-Poisson bracket to a supermanifold.
\section{Structure of $n$-ary $\Psi$-bracket in superspace $\mathbb R^{n|1}$}
In this section we study the structure and properties of $n$-ary $\Psi$-bracket defined in the previous section in the simplest case of the superspace $\mathbb R^{n|1}$, i.e. in the case of the superspace with $n$ real coordinates $x^\mu$ and one Grassmann coordinate $\theta$.

\vskip.3cm
\noindent
Let $\mathbb R^{n|1}$ be the superspace with $n$ real coordinates $x^\mu$, one Grassmann coordinate $\theta$, and, as before, we denote the collection of real coordinates by $r$. Evidently in this case the structure of an even degree function is very simple, it depends only on real coordinates $x^\mu$ and does not depend on the Grassmann coordinate $\theta$, i.e. if $f\in C^\infty_0(\mathbb R^{n|1})$ then $f$ is a usual smooth function $f(r)$ of $n$ real variables $x^\mu$. Next the odd part of any diffeomorphism of this superspace consists only of one odd degree function $\psi(r,\theta)=\psi(r)\,\theta$, where $\psi(r)$ is usual real-valued function of $n$ real variables $x^\mu$, which satisfies $\psi(r)\neq 0$ at any point $r\in \mathbb R^n$. It is easy to find that the $n$-ary $\Psi$-bracket (\ref{definition of Nambu-Poisson n-bracket}) for $n$ even degree functions $f^1(r),f^2(r),\ldots,f^n(r)\in C^\infty_0(\mathbb R^{n|1})$ can be written in the form
\begin{equation}
\{f^1(r),f^2(r),\ldots,f^n(r)\}_{\Psi}=\frac{1}{\psi(r)}\,\{f^1(r),f^2(r),\ldots,f^n(r)\},
\label{Nambu-Poisson bracket for m=1}
\end{equation}
where at the right-hand side of the above formula we have the usual Nambu-Poisson bracket of functions $f^1(r),f^2(r),\ldots,f^n(r)$. Indeed the $n\times m$-matrix $\frak B_{12}$ is the zero matrix because any even degree function $f^\mu(r)$ does not depend on $\theta$. Hence the product $\frak B_{12}\,\frak B_{22}^{-1}\,\frak B_{21}$ vanishes, and
$\frak B_{22}=(\partial_{\theta}\psi(r,\theta))=(\psi(r))$.

\vskip.3cm
\noindent
Since the usual Nambu-Poisson $n$-ary bracket satisfies the Leibniz rule, it is easy to show that the $n$-ary $\Psi$-bracket (\ref{Nambu-Poisson bracket for m=1}) also satisfies the Leibniz rule. Indeed we have
\begin{eqnarray}
\{g\,h,f^1\ldots,f^{n-1}\}_\Psi \!\!&=&\!\! \frac{1}{\psi}\,\{g\,h,f^1,\ldots,f^{n-1}\}\nonumber\\
                     \!\!&=&\!\!\frac{1}{\psi}\big(g\,\{h,f^1,\ldots,f^{n-1}\}+h\,\{g,f^1,\ldots,f^{n-1}\}\big)\nonumber\\
    \!\!&=&\!\!g\,\{h,f^1\ldots,f^{n-1}\}_\Psi+h\,\{g,f^1\ldots,f^{n-1}\}_\Psi.\nonumber
\end{eqnarray}
In order to show that the $\Psi$-bracket (\ref{Nambu-Poisson bracket for m=1}) satisfies the Filippov-Jacobi identity we will use the identity
\begin{eqnarray}
\{f^1,\ldots,f^{n-1},g\}\!&&\!\{f^n,\ldots,f^{2n-1}\}\nonumber\\
           && = \sum_{i=0}^{2n-1}\,\{f^1,\ldots,f^{n-1},f^{n+i}\}\,\{f^n,\ldots,\underset{(i+1)}{g},\ldots,f^{2n-1}\}.
\label{identity I}
\end{eqnarray}
Now the Filippov-Jacobi identity for the $n$-ary $\Psi$-bracket can be proved by means of the Leibniz rule and the identity (\ref{identity I}).
\begin{prop}
The $n$-ary $\Psi$-bracket (\ref{definition of Nambu-Poisson n-bracket}) in the case of the superspace $\mathbb R^{n|1}$ takes on the form (\ref{Nambu-Poisson bracket for m=1}) and it is the Nambu-Poisson $n$-bracket, i.e. it satisfies the property of skew-symmetry, the Leibniz rule for a product of two functions and the Filippov-Jacoby identity.
\end{prop}
\section{Structure of $n$-ary $\Psi$-bracket in superspace $\mathbb R^{n|2}$}
The aim of this section is to study a structure and properties of the $n$-ary $\Psi$-bracket defined by means of the superdeterminant in the case of the real superspace with $n$ real coordinates $x^\mu$ and two Grassmann coordinates $\theta,\eta$. We find that the $n$-ary $\Psi$-bracket can be expressed as the sum of two $n$-ary brackets, where one bracket is the usual Nambu-Poisson bracket of smooth functions and the second is a new $n$-ary bracket, which depends on the collection of odd degree smooth functions, and this bracket is referred to as the $\chi$-bracket. We prove two theorems, which state that $n$-ary $\chi$-bracket as well as the whole $n$-ary $\Psi$-bracket are $n$-ary Nambu-Poisson brackets.

\vskip.3cm
\noindent
Any even degree smooth function $f(r,\theta,\eta)$ can be expanded in terms of Grassmann variables as follows
$$
f(r,\theta,\eta)=f_0(r)+f_1(r)\,\theta\eta,
$$
where $f_0(r),f_1(r)$ are smooth functions of $n$ variables $x^\mu$. For an even degree function $f$ we define $\pi(f)=f_0,\, \delta(f)=f-\pi(f)=f_1\,\theta\eta$. Let $\phi(r,\theta,\eta),\,\psi(r,\theta,\eta)$ be two odd degree smooth functions. Then we can expand them in terms of Grassmann coordinates as follows
$$
\phi(r,\theta,\eta)=\phi_1(r)\,\theta+\phi_2(r)\,\eta,\;\;
      \psi(r,\theta,\eta)=\psi_1(r)\,\theta+\psi_2(r)\,\eta,
$$
where $\phi_1(r),\phi_2(r),\psi_1(r),\psi_2(r)$ are smooth functions. We assume that the second order matrix
\begin{equation}
\frak B_{22}=\left(
               \begin{array}{cc}
                 \partial_\theta \phi & \partial_\eta \phi \\
                 \partial_\theta \psi & \partial_\eta \psi \\
               \end{array}
             \right)=\left(
                       \begin{array}{cc}
                         \phi_1 & \phi_2 \\
                         \psi_1 & \psi_2 \\
                       \end{array}
                     \right)
\end{equation}
is invertible, i.e. $\mbox{Det}\,\frak B_{22}\neq 0$. It is worth to mention that the determinant of this matrix plays an important role and will appear in the expressions for $\Psi$-bracket. The determinant of matrix $\frak B_{22}$ will be denoted by $\Delta$, i.e. $\Delta=\mbox{Det}\,\frak B_{22}$.

\vskip.3cm
\noindent
As before, the collection of odd degree functions $\{\phi(r,\theta,\eta), \psi(r,\theta,\eta)\}$, which plays the role of odd degree part of a diffeomorphism of the superspace $\mathbb R^{n|2}$, will be denoted by $\Psi$. The product of odd degree functions $\phi(r,\theta,\eta), \psi(r,\theta,\eta)$ is the even degree function, which can be expressed in terms of Grassmann coordinates $\theta,\eta$ and the determinant $\Delta$ of matrix $\frak B_{22}$ as follows
$$
\phi(r,\theta,\eta)\cdot \psi(r,\theta,\eta)=\Delta\,\theta\eta.
$$
The product $\phi(r,\theta,\eta)\cdot \psi(r,\theta,\eta)$ will be denoted by $\chi$, thus we can write $\chi=\Delta\,\theta\eta$.

\vskip.3cm
\noindent
Let $f^1,f^2,\ldots,f^n$ be even degree smooth functions defined on the superspace $\mathbb R^{n|2}$. Thus $f^\mu=f^\mu_0+f^\mu_1\,\theta\eta$. Now according to the definition of $n$-ary $\Psi$-bracket (\ref{definition of Nambu-Poisson n-bracket}) we have
\begin{equation}
\{f^1,f^2,\ldots,f^n\}_\Psi=
    \mbox{Sdet}\,\left(
     \begin{array}{ccccccc}
     \partial_{x^1}f^1 & \partial_{x^2}f^1 & \ldots & \partial_{x^n}f^1 & | & \partial_{\theta}f^1 & \partial_{\eta}f^1 \\
     \partial_{x^1}f^2 & \partial_{x^2}f^2 & \ldots & \partial_{x^n}f^2 & | & \partial_{\theta}f^2 & \partial_{\eta}f^2 \\
     \ldots & \ldots & \ldots & \ldots & | & \ldots & \ldots \\
     \partial_{x^1}f^n & \partial_{x^2}f^n & \ldots & \partial_{x^n}f^n & | & \partial_{\theta}f^n & \partial_{\eta}f^n \\
     -- & -- & -- & -- & -- & -- & --  \\
     \partial_{x^1}\phi & \partial_{x^2}\phi & \ldots & \partial_{x^n}\phi & | & \partial_{\theta}\phi & \partial_{\eta}\phi \\
     \partial_{x^1}\psi & \partial_{x^2}\psi & \ldots & \partial_{x^n}\psi & | & \partial_{\theta}\psi & \partial_{\eta}\psi \\
     \end{array}
     \right),
\label{n-ary bracket for m=2}
\end{equation}
where dotted lines split the supermatrix into the even and odd degree blocks. In order to compute the superdeterminant we have to compute the usual determinant of the following matrix
\begin{eqnarray}
\frak A \!\!\!&=&\!\!\!\left(\begin{array}{cccc}
                              \partial_{x^1}f^1 & \partial_{x^2}f^1 & \ldots & \partial_{x^n}f^1 \\
                              \partial_{x^1}f^2 & \partial_{x^2}f^2 & \ldots & \partial_{x^n}f^2 \\
                              \ldots & \ldots & \ldots & \ldots  \\
                              \partial_{x^1}f^n & \partial_{x^2}f^n & \ldots & \partial_{x^n}f^n\\
                              \end{array}
                            \right)-\left(
                                      \begin{array}{cc}
                                        \partial_{\theta}f^1 & \partial_{\eta}f^1\\
                                        \partial_{\theta}f^2 & \partial_{\eta}f^2\\
                                        \ldots               &\ldots             \\
                                        \partial_{\theta}f^n & \partial_{\eta}f^n\\
                                        \end{array}
                                    \right)\nonumber\\
                                    &&\qquad\qquad\qquad\qquad\qquad\times\left(
                                     \begin{array}{cc}
                                       \partial_{\theta}\phi & \partial_{\eta}\phi \\
                                       \partial_{\theta}\psi & \partial_{\eta}\psi \\
                                     \end{array}
                                   \right)^{-1}\left(
                                                 \begin{array}{cccc}
                                                   \partial_{x^1}\phi & \partial_{x^2}\phi & \ldots & \partial_{x^n}\phi \\
                                                   \partial_{x^1}\psi & \partial_{x^2}\psi & \ldots & \partial_{x^n}\psi \\
                                                 \end{array}
                                               \right).\nonumber
\end{eqnarray}
If an element of this $n$th order matrix is denoted by $\frak a^\mu_\nu$ then we find
\begin{equation}
\frak a^\mu_\nu=\frac{\partial f^\mu}{\partial x^\nu}+\Delta^{-1}\frac{\partial \Delta}{\partial x^\nu}\;\delta(f^\mu).
\end{equation}
Making use of this formula for an element of the matrix $\frak A$ we prove the following statement
\begin{prop}
The $n$-ary $\Psi$-bracket (defined in (\ref{n-ary bracket for m=2})) for $n$ even degree smooth functions $f^\mu$, where $f^\mu=f^\mu_0+f^\mu_1\,\theta\eta$, can be expressed as follows
\begin{equation}
\{f^1,f^2,\ldots,f^n\}_\Psi=
       \frac{1}{\Delta}\{f^1,f^2,\ldots,f^n\}+\frac{1}{\Delta}\,\{f^1,f^2,\ldots,f^n\}_{\chi},
\label{n-ary bracket as the sum of Nambu bracket and new bracket}
\end{equation}
where $\Delta$ is the determinant of matrix $\frak B_{22}$, $\{f^1,f^2,\ldots,f^n\}$ is the usual Nambu-Poisson $n$-bracket
$$
\{f^1,f^2,\ldots,f^n\}=\mbox{Det}\,\left(
                              \begin{array}{cccc}
                              \partial_{x^1}f^1 & \partial_{x^2}f^1 & \ldots & \partial_{x^n}f^1 \\
                              \partial_{x^1}f^2 & \partial_{x^2}f^2 & \ldots & \partial_{x^n}f^2 \\
                              \ldots & \ldots & \ldots & \ldots  \\
                              \partial_{x^1}f^n & \partial_{x^2}f^n & \ldots & \partial_{x^n}f^n\\
                              \end{array}
                            \right),
$$
$\partial_{x^\nu}f^\mu=\partial_{x^\nu}f^\mu_0+\partial_{x^\nu}f^\mu_1\cdot\theta\eta$, and
\begin{equation}
\{f^1,f^2,\ldots,f^n\}_{\chi}=\frac{1}{\Delta}\sum_{\mu=1}^n\,\delta(f^\mu)\cdot\{f^1,f^2,\ldots,\underset{(\mu)}{\Delta},\ldots, f^n\},
\label{formula I for chi bracket}
\end{equation}
where $\chi=\phi\cdot\psi=\Delta\,\theta\eta$ is the even degree smooth function.
\label{proposition which gives formula for n-ary bracket}
\end{prop}
Proposition (\ref{proposition which gives formula for n-ary bracket}) shows that the $n$-ary $\Psi$-bracket (\ref{definition of Nambu-Poisson n-bracket}) defined with the help of superdeterminant can be considered as the extension of the usual Nambu-Poisson $n$-bracket by means of the new $n$-ary bracket $\{f^1,f^2,\ldots,f^n\}_\chi$ defined in (\ref{formula I for chi bracket}). Obviously this new $n$-ary bracket is totally skew-symmetric. We can write this new bracket with the help of Berezin integral over Grassmann algebra generated by coordinates $\theta,\eta$. We remind the definition of this integral in particular case of two Grassmann variables. The Berezin integral is defined by
\begin{equation}
\int\,d\theta=\int\,d\eta=0,\;\;\;\int\,\theta\,d\theta=\int\,\eta\,d\eta=1,
\end{equation}
and a multiple integral is computed as iterated integral. Hence for even degree smooth function $f^\mu(r,\theta,\eta)=f^\mu_0(r)+f^\mu_1(r)\,\theta\eta$ we have
$$
\int\,f^\mu(r,\theta,\eta)\,d\eta\, d\theta=f^\mu_1(r).
$$
Now the formula for $n$-ary $\chi$-bracket (\ref{formula I for chi bracket}) can be written by means of Berezin integral as follows
\begin{equation}
\{f^1,f^2,\ldots,f^n\}_{\chi}=\frac{1}{\Delta}\sum_{\mu=1}^n\;\int f^\mu\,d\eta\, d\theta\cdot\{f^1,f^2,\ldots,\underset{(\mu)}{\chi},\ldots, f^n\},
\label{formula II for n-ary chi-bracket}
\end{equation}
Particularly if our superspace is the super plane $\mathbb R^{2|2}$ with two real coordinates $r=(x,y)$ and two Grassmann coordinates $\theta,\eta$ then the $n$-ary $\Psi$-bracket (\ref{n-ary bracket as the sum of Nambu bracket and new bracket}) gives the binary bracket of two even degree smooth functions $f,g$, which can be written as
\begin{equation}
\{f,g\}_\Psi=\frac{1}{\Delta}\,\{f,g\}+\frac{1}{\Delta}\,\{f,g\}_\chi,
\label{binary Psi bracket}
\end{equation}
where the first bracket at the right hand side is the usual Poisson bracket for even degree superfunctions
\begin{eqnarray}
\{f,g\}=\{\pi(f),\pi(g)\}+\{\pi(f),\delta(g)\}+\{\delta(f),\pi(g)\}.\nonumber
\end{eqnarray}
The binary $\chi$-bracket in (\ref{binary Psi bracket}) can be written in the form
\begin{equation}
\{f,g\}_\chi=\frac{1}{\Delta}\big(\delta(f)\,\{\Delta,\pi(g)\}+\delta(g)\,\{\pi(f),\Delta\}\big).
\end{equation}
In the superspace $\mathbb R^{3|2}$ with three real coordinates $r=(x,y,z)$ and two anti-commutating coordinates $\theta,\eta$ the $n$-ary $\Psi$-bracket (\ref{n-ary bracket as the sum of Nambu bracket and new bracket}) yields the triple bracket of three even degree smooth functions $f,g,h$,
which can be written as
$$
\{f,g,h\}_\Psi=\frac{1}{\Delta}\,\{f,g,h\}+\frac{1}{\Delta}\,\{f,g,h\}_\chi,
$$
where
\begin{eqnarray}
\{f,g,h\} = \{\pi(f),\pi(g),\pi(h)\}+\{\delta(f),\pi(g),\pi(h)\}
               &+&\{\pi(f),\delta(g),\pi(h)\}\nonumber\\
                     &&\;\;\;+\{\pi(f),\pi(g),\delta(h)\}\big),\nonumber
\end{eqnarray}
and
\begin{eqnarray}
\{f,g,h\}_\chi = \frac{1}{\Delta}\big(\delta(f)\,\{\Delta,g,h\}+\delta(g)\,\{f,\Delta,h\}+\delta(h)\,\{f,g,\Delta\}\big)\nonumber.
\end{eqnarray}
We begin the study of the properties of the whole $n$-ary $\Psi$-bracket (\ref{n-ary bracket as the sum of Nambu bracket and new bracket}) with the study of the properties of the $n$-ary $\chi$-bracket, which stands in the second place in (\ref{n-ary bracket as the sum of Nambu bracket and new bracket}), that is, the $n$-ary $\chi$-bracket that extends the $n$-ary Nambu-Poisson bracket to the $n$-ary $\Psi$-bracket.
\begin{thm}
The $n$-ary $\chi$-bracket of even degree smooth functions defined in (\ref{formula I for chi bracket}) satisfies:
\begin{enumerate}
\item The property of totally skew-symmetry, which means that any permutation of arguments of $n$-ary $\chi$-bracket changes its sign according to the parity of permutation;
\item The Leibniz rule for a product of functions, i.e.
$$
\{g\cdot h,f^1,f^2,\ldots,f^{n-1}\}_\chi=g\,\{h,f^1,f^2,\ldots,f^{n-1}\}_\chi+
                      h\,\{g,f^1,f^2,\ldots,f^{n-1}\}_\chi;
$$
\item The Filippov-Jacobi identity (fundamental identity), i.e.
\begin{eqnarray}
\{f^1,&f^2&,\ldots,f^{n-1},\{g^1,g^2,\ldots,g^n\}_\chi\}_\chi\nonumber\\
    \!\!&=&\!\!\sum_{\mu=1}^n\,\{g^1,g^2,\ldots,g^{\mu-1},\{f^1,f^2,\ldots,f^{n-1},g^\mu\}_\chi,g^{\mu+1},\ldots,g^n\}_\chi.
\label{Filippov-Jacobi identity for chi-bracket}
\end{eqnarray}
\end{enumerate}
\label{theorem about properties of chi-bracket}
\end{thm}
The property of totally skew-symmetry of $n$-ary $\chi$-bracket follows immediately from the definition (\ref{formula I for chi bracket}). Let us show that the $n$-ary $\chi$-bracket satisfies the Leibniz rule. Applying the definition of $\chi$-bracket we get
\begin{eqnarray}
\{g\,h,f^1,f^2,\ldots,f^{n-1}\}_\chi \!\!&=&\!\!
   \frac{1}{\Delta}\Big((gh)_1\,\{\chi,f^1,f^2,\ldots,f^{n-1}\}\nonumber\\
     &&+ f^1_1\,\{gh,\chi,f^2,\ldots,f^{n-1}\}+
                     \ldots \nonumber\\
                     && + f^{n-1}_1\,\{gh,f^1,\ldots,f^{n-2},\chi\}\Big).\nonumber
\end{eqnarray}
It is easily verified that the differentiation property (Leibniz rule) of the Nambu-Poisson $n$-ary bracket extends to even degree superfunctions. Now we use this differentiation property of the Nambu-Poisson $n$-ary bracket in the case of even degree functions and obtain
\begin{eqnarray}
\{g\,h,f^1,f^2,\ldots,f^{n-1}\}_\chi \!\!\!&=&\!\!\!\frac{1}{\Delta}\Big((gh)_1\,\{\chi,f^1,f^2,\ldots,f^{n-1}\}+
                                   f^1_1\,g\;\{h,\chi,f^2,\ldots,f^{n-1}\}\nonumber\\
          &&\quad +f^1_1\,h\;\{g,\chi,f^2,\ldots,f^{n-1}\}+\ldots\nonumber\\
           &&\quad + f^{n-1}_1\,g\;\{h,f^1,\ldots,f^{n-2},\chi\}\nonumber\\
          &&\quad +f^{n-1}_1\,h\;\{g,f^1,\ldots,f^{n-2},\chi\}\Big).\nonumber
\end{eqnarray}
Now we notice that the coefficient function of $\theta\eta$ in the product $g\,h$ at the right hand side of the previous formula can be written in the form $g_1\,h+g\,h_1$, that is, $(g\,h)_1=g_1\,h+g\,h_1$. Indeed $(g\,h)_1=g_0\,h_1+g_1\,h_0$, but we can add the term $g_1\,\theta\eta$ to $g_0$ (making it $g$) and analogously $h_1\,\theta\eta$ to $h_0$ (making it $h$) because these additional terms will be \textquote{killed} by the second factor of the product $\{\chi,f^1,f^2,\ldots,f^{n-1}\}$, where $\chi=\Delta\,\theta\eta$.
Thus
\begin{eqnarray}
\{g\,h,f^1,f^2,\ldots,f^{n-1}\}_\chi \!\!\!&=&\!\!\!g\,\{h,f^1,f^2,\ldots,f^{n-1}\}_\chi+h\,\{g,f^1,f^2,\ldots,f^{n-1}\}_\chi.
\nonumber
\end{eqnarray}
Now our aim is to prove the Filippov-Jacobi identity (\ref{Filippov-Jacobi identity for chi-bracket}). First we will simplify the left hand side of this identity. We remind that any even degree function $g$ can be expanded in Grassmann variables $\theta,\eta$ as follows $g_0(r)+g_1(r)\,\theta\eta$, and the first function of this expansion $g_0$ will be referred to as 0-component of an even degree function $g$ and $g_1(r)$ as 1-component of a function $g$. Now we see that the formula for $\chi$-bracket (\ref{formula I for chi bracket}) can be written in the form
$$
\{f^1,f^2,\ldots,f^n\}_{\chi}=\frac{1}{\Delta}\sum_{\mu=1}^n\,f^\mu_1\cdot\{f^1_0,f^2_0,\ldots,\underset{(\mu)}{\Delta},\ldots, f^n_0\}\cdot\theta\eta,
$$
which clearly shows that, first, $\chi$-bracket has no 0-component, which is zero, and, second, in order to calculate a $\chi$-bracket we need to use the 0-component of each function $f^\mu$ in Nambu-Poisson bracket and to use its 1-component when function is taken out of Nambu-Poisson bracket and replaced by $\Delta$. Consequently when we calculate the outer $\chi$-bracket of the double $\chi$-bracket at the left hand side of (\ref{Filippov-Jacobi identity for chi-bracket}) the first $n-1$ terms in our sum vanish because the last argument of $n$-ary Nambu-Poisson bracket is the 0-component of $\{g^1,g^2,\ldots,g^n\}_\chi$, but this is zero. Hence we have only one non-trivial term in the expansion of outer $\chi$-bracket at the left hand side of (\ref{Filippov-Jacobi identity for chi-bracket}), which can be written as
\begin{equation}
\frac{1}{\Delta}\big(\{g^1,g^2,\ldots,g^n\}_\chi\big)_1\;\{f^1_0,f^2_0,\ldots,f^{n-1}_0,\Delta\}\cdot\theta\eta,
\label{proof of theorem for chi-bracket 2}
\end{equation}
where $\big(\{g^1,g^2,\ldots,g^n\}_\chi\big)_1$ denotes the 1-component of $\chi$-bracket. Analogously the right hand side of the Filippov-Jacobi identity can be written as
\begin{eqnarray}
\frac{1}{\Delta}\sum_\mu\,\Big(\{f^1,f^2,\ldots,f^{n-1},g^\mu\}_\chi\Big)_1\;\{g_0^1,\ldots,\underset{(\mu)}{\Delta},\ldots,g^n_0\}\cdot\theta\eta.
\label{proof of theorem for chi-bracket 1}
\end{eqnarray}
The 1-component of the $\chi$-bracket of the previous formula can be written as
\begin{eqnarray}
\Big(\{f^1,f^2,\ldots,f^{n-1},g^\mu\}_\chi\Big)_1 \!\!\!&=&\!\!\!\frac{1}{\Delta}\sum_{\nu=1}^{n-1}\,f_1^\nu\,\{f^1_0,\ldots,\underset{(\nu)}{\Delta},\ldots,f^{n-1}_0,g^\mu_0\}\nonumber\\
          &&\qquad\qquad\qquad + \frac{g^\mu_1}{\Delta}\,\{f^1_0,\ldots,\ldots,f^{n-1}_0,\Delta\}.\nonumber
\end{eqnarray}
If we substitute the right hand side of this formula into the formula (\ref{proof of theorem for chi-bracket 1}) and take only the last term then we get the expression
\begin{eqnarray}
&&\frac{1}{\Delta}\{f^1_0,\ldots,\ldots,f^{n-1}_0,\Delta\}\sum_{\mu}\,\frac{g^\mu_1}{\Delta}\{g^1_0,\ldots,\underset{(\mu)}{\Delta},\ldots,g^n_0\}=\nonumber\\
   &&\qquad\qquad\qquad\qquad\qquad\frac{1}{\Delta}\{f^1_0,\ldots,\ldots,f^{n-1}_0,\Delta\}\{g^1,g^2,\ldots,g^n\}_\chi,\nonumber
\end{eqnarray}
which is exactly the left hand side of the Filippov-Jacobi identity written in the form (\ref{proof of theorem for chi-bracket 2}). Thus in order to complete the proof we need to show that the sum of remained terms is zero, i.e. the following double sum
\begin{equation}
\frac{1}{\Delta^2}\sum_{\mu=1}^n\,\;\sum_{\nu=1}^{n-1}\,f_1^\nu\,\{f^1_0,\ldots,\underset{(\nu)}{\Delta},\ldots,f^{n-1}_0,g^\mu_0\}\;
                    \{g_0^1,\ldots,\underset{(\mu)}{\Delta},\ldots,g^n_0\}\cdot\theta\eta
                    \label{proof of theorem for chi-bracket 3}
\end{equation}
vanishes. If we rearrange the order of summation and first consider the sum
$$
\sum_{\mu=1}^n\,\{f^1_0,\ldots,\underset{(\nu)}{\Delta},\ldots,f^{n-1}_0,g^\mu_0\}\;
                    \{g_0^1,\ldots,\underset{(\mu)}{\Delta},\ldots,g^n_0\},
$$
where $\nu$ is fixed, then we can write this sum by means of the identity (\ref{identity I}) in the form
$$
\{f^1_0,\ldots,\underset{(\nu)}{\Delta},\ldots,f^{n-1}_0,\Delta\}.
$$
Hence we get the $n$-ary Nambu-Poisson bracket with two equal arguments ($\Delta$) and, because of total skew-symmetry of $n$-ary Nambu-Poisson bracket, this $n$-ary Nambu-Poisson bracket vanishes. Since this holds for every value of integer $\nu=1,2\ldots,n-1$, the whole sum (\ref{proof of theorem for chi-bracket 3}) is zero and this ends the proof.

\vskip.3cm
\noindent
An important consequence of this proved theorem is that the algebra of even degree functions $C^\infty_0(\mathbb R^{n|2})$ defined on the super space $\mathbb R^{n|2}$ and equipped with the $n$-ary $\chi$-bracket (\ref{formula I for chi bracket}) is the $n$-Lie algebra. We can prove a similar theorem for the whole $n$-ary $\Psi$-bracket (\ref{n-ary bracket for m=2}).
\begin{thm}
The $n$-ary $\Psi$-bracket of even degree smooth functions defined in (\ref{n-ary bracket for m=2}) satisfies:
\begin{enumerate}
\item The property of totally skew-symmetry, which means that any permutation of arguments of $n$-ary $\Psi$-bracket changes its sign according to the parity of permutation;
\item The Leibniz rule for a product of functions, i.e.
$$
\{g\cdot h,f^1,f^2,\ldots,f^{n-1}\}_\Psi=g\,\{h,f^1,f^2,\ldots,f^{n-1}\}_\Psi+
                      h\,\{g,f^1,f^2,\ldots,f^{n-1}\}_\Psi;
$$
\item The Filippov-Jacobi identity (fundamental identity), i.e.
\begin{eqnarray}
\{f^1,f^2,\!\!\!&\ldots&\!\!\!,f^{n-1},\{g^1,g^2,\ldots,g^n\}_\Psi\}_\Psi\nonumber\\
    \!\!&=&\!\!\sum_{\mu=1}^n\,\{g^1,g^2,\ldots,g^{\mu-1},\{f^1,f^2,\ldots,f^{n-1},g^\mu\}_\Psi,g^{\mu+1},\ldots,g^n\}_\Psi.
\label{Filippov-Jacobi identity for chi-bracket}
\end{eqnarray}
\end{enumerate}
\label{theorem about properties of chi-bracket}
\end{thm}
In order to prove this theorem we will use Proposition \ref{proposition which gives formula for n-ary bracket} and Theorem \ref{theorem about properties of chi-bracket}. Particularly Proposition \ref{proposition which gives formula for n-ary bracket} gives the explicit formula for $n$-ary $\Psi$-bracket (\ref{n-ary bracket as the sum of Nambu bracket and new bracket}), which shows that $\Psi$-bracket is the sum of $n$-ary Nambu-Poisson bracket and $n$-ary $\chi$-bracket. Thus the property of total skew-symmetry and the Leibniz rule (differentiation property) for $n$-ary $\Psi$-bracket follows immediately from this formula and the properties of $\chi$-bracket proved in Theorem \ref{theorem about properties of chi-bracket}. Thus the theorem will be proved if we show that $n$-ary $\Psi$-bracket satisfies the Filippov-Jacobi identity.

\vskip.3cm
\noindent
Making use of the formula (\ref{n-ary bracket as the sum of Nambu bracket and new bracket}), we can write the left hand side of the Filippov-Jacobi identity for $n$-ary $\Psi$-bracket as follows
\begin{eqnarray}
&&\frac{1}{\Delta}\,\{f^1,\ldots,f^{n-1},\frac{1}{\Delta}\,\{g^1,\ldots,g^n\}\}+
                       \frac{1}{\Delta}\,\{f^1,\ldots,f^{n-1},\frac{1}{\Delta}\,\{g^1,\ldots,g^n\}_\chi\}\nonumber\\
    &&\quad\quad +\frac{1}{\Delta}\,\{f^1,\ldots,f^{n-1},\frac{1}{\Delta}\,\{g^1,\ldots,g^n\}\}_\chi+
                           \frac{1}{\Delta}\,\{f^1,\ldots,f^{n-1},\frac{1}{\Delta}\,\{g^1,\ldots,g^n\}_\chi\}_\chi.\nonumber
\end{eqnarray}
The first double $n$-ary bracket of this expression is no different from the double $n$-ary bracket generated with the help of the $n$-ary bracket (\ref{Nambu-Poisson bracket for m=1}). But it was proved in Section 3 that the $n$-ary bracket (\ref{Nambu-Poisson bracket for m=1}) satisfies the Filippov-Jacobi identity. Hence the first double bracket
$$
\frac{1}{\Delta}\,\{f^1,\ldots,f^{n-1},\frac{1}{\Delta}\,\{g^1,\ldots,g^n\}\},
$$
at the left hand side of Filippov-Jacobi identity is equal to the sum
$$
\sum_\mu\;\frac{1}{\Delta}\,\{g^1,\ldots,g^{\mu-1},\frac{1}{\Delta}\,\{f^1,\ldots,f^{n-1},g^\mu\},g^{\mu+1},\ldots,g^n\}
$$
at the right hand side of the Filippov-Jacobi identity. Analogously applying Theorem \ref{theorem about properties of chi-bracket} we can conclude that the last double $n$-ary $\chi$-bracket
$$
\frac{1}{\Delta}\,\{f^1,\ldots,f^{n-1},\frac{1}{\Delta}\,\{g^1,\ldots,g^n\}_\chi\}_\chi
$$
at the left hand side of the Filippov-Jacobi identity is equal to
$$
\sum_\mu\;\frac{1}{\Delta}\,\{g^1,\ldots,g^{\mu-1},\frac{1}{\Delta}\,\{f^1,\ldots,f^{n-1},g^\mu\}_\chi,g^{\mu+1},\ldots,g^n\}_\chi
$$
at the right hand side of the Filippov-Jacobi identity. Thus it remains to prove the identity
\begin{eqnarray}
&&\frac{1}{\Delta}\,\{f^1,\ldots,f^{n-1},\frac{1}{\Delta}\,\{g^1,\ldots,g^n\}_\chi\}+
          \frac{1}{\Delta}\,\{f^1,\ldots,f^{n-1},\frac{1}{\Delta}\,\{g^1,\ldots,g^n\}\}_\chi\nonumber\\
 &&\qquad =\sum_\mu\;\frac{1}{\Delta}\,\{g^1,\ldots,g^{\mu-1},\frac{1}{\Delta}\,\{f^1,\ldots,f^{n-1},g^\mu\}_\chi,g^{\mu+1},\ldots,g^n\}\nonumber\\
 &&\qquad\qquad+\sum_\mu\;\frac{1}{\Delta}\,\{g^1,\ldots,g^{\mu-1},\frac{1}{\Delta}\,\{f^1,\ldots,f^{n-1},g^\mu\},g^{\mu+1},\ldots,g^n\}_\chi.
\label{auxiliary identity in the proof of theorem 2}
\end{eqnarray}
The first double $n$-ary bracket at the left hand side ($\chi$-bracket inside the $n$-ary Nambu-Poisson bracket) can be written as follows
\begin{eqnarray}
&&\frac{1}{\Delta^3}\sum_\mu\,g^\mu_1\,\{f^1_0,\ldots,f^{n-1}_0,
                           \{g^1_0,\ldots,\underset{(\mu)}{\Delta},\ldots,g_0^n\}\}\,\theta\eta\nonumber\\
    &&\qquad\qquad\qquad+ \frac{1}{\Delta^3}\{f^1_0,\ldots,f^{n-1}_0,\Delta\}
                           \{g^1_0,\ldots,g_0^n\}\,\theta\eta\nonumber\\
    &&\qquad\qquad\qquad\qquad\qquad + \frac{1}{\Delta}\{f^1,\ldots,f^{n-1},\frac{1}{\Delta^2}\}
                           \{g^1,\ldots,g^n\}\}_\chi\,\theta\eta,\nonumber
\end{eqnarray}
and similarly the second term at the left hand side (Nambu-Poisson bracket inside the $\chi$-bracket) can be expressed as follows
\begin{eqnarray}
\frac{1}{\Delta^3}\sum_{\mu=1}^{n-1}\,f^\mu_1\,\{f^1_0,\ldots,\underset{(\mu)}{\Delta},\ldots,f^{n-1}_0,
                           \{g^1_0,&\ldots&,g_0^n\}\}\,\theta\eta\nonumber\\
    && + \frac{1}{\Delta^3}\{f^1_0,\ldots,f^{n-1}_0,\Delta\}
                           \{g^1,\ldots,g^n\}_1\,\theta\eta.\nonumber
\end{eqnarray}
Applying these formulae to mixed Nambu-Poisson and $\chi$-bracket at the right hand side we can verify the identity (\ref{auxiliary identity in the proof of theorem 2}) and this ends the proof of the theorem.

\section{Discussion}
The construction proposed in this paper, which is based on the notion of Berezinian, leads to an n-ary Nambu-Poisson bracket of even degree functions defined on a superspace $\mathbb R^{n|m}$. This $n$-ary Nambu-Poisson bracket determines the structure of the Nambu-Poisson algebra on the algebra of even degree functions. The structure of $n$-ary Nambu-Poisson bracket also includes an ordered set of odd degree functions $(\psi^1,\psi^2,\ldots,\psi^m)$, which are considered, in the framework of our approach, as parameters of Nambu-Poisson bracket. Hence we have the family of $n$-ary Nambu-Poisson brackets, where the parameter of the family is an ordered set of $m$ odd degree functions, which determines the diffeomorphism of the odd degree part of a superspace $\mathbb R^{n|m}$. It is important here that the dependence of $n$-ary Nambu-Poisson bracket on odd degree functions is skew-symmetric, which follows from the properties of superdeterminant. Hence if we subject odd degree functions to a permutation, then the sign of $n$-ary Nambu-Poisson bracket changes according to the parity of a permutation. Thus we have
$$
\{f^1,f^2,\ldots,f^n\}_{(\psi^{\sigma(1)},\psi^{\sigma(2)},\ldots,\psi^{\sigma(m)})}=
        (-1)^{|\sigma|}\{f^1,f^2,\ldots,f^n\}_{(\psi^1,\psi^2,\ldots,\psi^m)},
$$
where $\sigma$ is a permutation of integers and $|\sigma|$ is the parity of this permutation. In this our construction of the $n$-ary Nambu-Poisson bracket differs from the construction proposed in the paper \cite{Sasakibara}. First, the bracket proposed in \cite{Sasakibara} is $(n+m)$-ary bracket in which odd degree functions enter as arguments, that is, on par with even degree functions. Second, the bracket is graded skew-symmetric, which particularly means that if we subject odd degree functions in this bracket to a permutation then the sign of the bracket does not change. Hence we can say that the family of $n$-ary Nambu-Poisson brackets constructed and studied in this paper and the super Nambu-Poisson bracket, which is proposed and studied in \cite{Sasakibara}, have different structures, which lead to different algebraic structures on superspace.

\subsection*{Acknowledgment}
The author gratefully acknowledges that this work was financially supported by the institutional funding IUT20-57 of the Estonian Ministry of Education and Research.

\end{document}